\documentclass[12pt]{article}

\usepackage[utf8x]{inputenc}
\usepackage{amsfonts}
\usepackage[]{hyperref}
\hypersetup{linktocpage}
\usepackage{mathtools}
\usepackage{amsmath,amssymb,latexsym,cite}
\usepackage{tikz-cd}

\begin{document}

\begin{center}

\vspace*{0.5in}

{\large\bf Landau-Ginzburg models for\\ certain fiber products with curves}

\vspace{0.2in}

Zhuo Chen$^1$, Tony Pantev$^2$, Eric Sharpe$^3$

\vspace*{0.2in}

\begin{tabular}{cc}
{ \begin{tabular}{c}
$^{1,3}$ Dep't of Physics\\
Virginia Tech\\
850 West Campus Dr.\\
Blacksburg, VA  24061
\end{tabular} } &
{ \begin{tabular}{c}
$^2$ Dep't of Mathematics \\
University of Pennsylvania \\
David Rittenhouse Lab. \\
209 South 33rd Street \\
Philadelphia, PA 19104-6395
\end{tabular} }
\end{tabular}

{\tt zhuo2012@vt.edu}, {\tt tpantev@math.upenn.edu}, {\tt ersharpe@vt.edu}

$\,$

\end{center}

In this paper we describe a physical realization of a
family of non-compact K{\"a}hler 
threefolds with trivial canonical bundle
in hybrid Landau-Ginzburg models,
motivated by some recent non-K\"ahler solutions of Strominger systems, 
and utilizing
some recent ideas from GLSMs.  We consider
threefolds given as fiber products of compact genus $g$ 
Riemann surfaces and noncompact threefolds.
Each genus $g$ Riemann surface is constructed using recent GLSM tricks,
as a double cover of $\mathbb{P}^1$ 
branched over a degree $2g + 2$ locus, realized via nonperturbative
effects rather than as the critical locus of a superpotential. 
We focus in particular on special cases corresponding to a set of
K\"ahler twistor spaces of certain hyperK\"ahler four-manifolds,
specifically
the twistor spaces of $\mathbb{R}^4$, $\mathbb{C}^2 / \mathbb{Z}_k$,
and $S^1 \times \mathbb{R}^3$. 
We check in all cases that the condition for trivial
canonical bundle arising
physically matches the mathematical constraint.

\begin{flushleft}
June 2018
\end{flushleft}

\newpage

\tableofcontents

\section{Introduction}

Over the years, there has been much work on non-K\"ahler solutions of heterotic
compactifications with $H$ flux, known as the
Strominger system \cite{Strominger:1986uh}.
This paper is inspired by the recent
work \cite{Fei:2017ctw} in which a new family of 
compact non-K{\" a}hler analogues of Calabi-Yau threefolds, a new set of
potential solutions to the Strominger system, was
constructed.   In this paper, we do not construct physical theories
for non-K\"ahler targets, but instead apply recent tricks in GLSMs to
build physical theories for K\"ahler analogues of the fiber products
discussed in \cite{Fei:2017ctw}.

Specifically,
suppose $M$ is a compact hyperK{\"a}hler manifold with real dimension four and 
$\Sigma$ is a compact Riemann surface of genus $g \geq 3$.
Then, a manifold $X$ with holomorphically-trivial canonical bundle 
can be constructed as the pullback
\begin{equation*}
\begin{tikzcd}
X= \varphi^* Z = Z \times_{\mathbb{P}^1} \Sigma \arrow{d} \arrow{r} & Z = M \times \mathbb{P}^1 \arrow[d, "\pi"] \\
\Sigma \arrow[r,"\varphi"] & \mathbb{P}^1
\end{tikzcd}
\end{equation*} 
where $Z$ is the twistor space of $M$ together with the natural holomorphic 
projection $\pi$ and $\varphi$ is a nonconstant holomorphic map. 
It was argued in \cite{Fei:2017ctw} that
the threefold $X$ has trivial canonical bundle as long as
\begin{equation*}
 \varphi^* \mathcal{O}(2) \cong K_{\Sigma}.
\end{equation*}
In this paper, the curve $\Sigma$ will be constructed as a branched
double cover of ${\mathbb P}^1$, for which case the condition above for the
fiber product to have trivial
canonical bundle reduces to simply $g=3$, independent of the details of $M$.

Motivated by the mathematical
construction above, we will give a physical realization of
threefolds of trivial canonical bundle constructed as fiber products of
genus $g$ curves with noncompact K\"ahler threefolds, including
as special cases certain noncompact K\"ahler\footnote{
Most twistor spaces are not K\"ahler.
} twistor spaces.
We will take the Riemann surface of genus $g$ and the holomorphic map 
$\varphi$ to be a branched double cover over $\mathbb{P}^1$, realized
physically via nonperturbative tricks as in \cite{Caldararu:2007tc}. 
Because we describe fiber products with K\"ahler threefolds,
including K\"ahler twistor spaces of certain hyperK\"ahler four-manifolds,
the Calabi-Yau threefolds we realize are 
non-compact and K{\"a}hler, as opposed to non-K\"ahler
spaces of trivial canonical bundle which were the focus of \cite{Fei:2017ctw}. 

We will construct higher-energy theories that realize these geometries as
(2,2) supersymmetric
hybrid Landau-Ginzburg models.  These hybrid models do not seem to have
a UV description as GLSMs, though some GLSMs do come close, as we shall
explain later.

We begin in section~\ref{sec: map} with a review of GLSMs for genus
$g$ curves, constructed via nonperturbative methods as branched double
covers.  
In section~\ref{sect:genl-case} we construct (2,2) supersymmetric hybrid 
Landau-Ginzburg models for the fiber products above, of curves with
a few noncompact K\"ahler threefolds.
In section~\ref{sec:hybrid} we specialize to fiber products of curves
and twistor spaces, which arise as special cases.
In appendix~\ref{sec:math} we review some pertinent mathematics.

Although we are not able to give physical realizations of any
non-K\"ahler geometries in this paper, it is our hope that the ideas
we present here will later be extended to non-K\"ahler fiber product
constructions.

Finally, before starting, we should add a caution.
We discuss non-compact K\"ahler manifolds with trivial canonical
bundle.  However, Yau's theorem does not apply to non-compact cases,
so
it is possible\footnote{
One set of examples is discussed in \cite{gyz}.  A second set of examples
arises by removing an anticanonical divisor with deep singularities from 
a projective manifold, then the complement will not have a complete Ricci-flat
metric, though it may still have a non-complete Ricci-flat metric.
A third set of examples arises from the cotangent bundle of a compact
K\"ahler manifold $X$.  On a tubular neighborhood of the zero section,
one can find a unique Ricci-flat metric which restricts to the given
K\"ahler metric on the zero section and is circle invariant.  This metric 
extends to a Ricci-flat metric on the whole cotangent bundle if and only
if $X$ is a homogeneous Fano or semi-abelian variety.  For example, if
$X$ is a Riemann surface of genus at least two, there is a Ricci-flat metric
on a disk bundle of fixed radius but it cannot be extended beyond that bundle
to the entire cotangent bundle.
} that some
might not have Ricci-flat metrics.  
(Nevertheless, we will sometimes call these noncompact K\"ahler spaces of
trivial canonical bundle, ``Calabi-Yau's,'' though this terminology is
inaccurate.)
This is an issue for both the spaces
themselves as well as for Landau-Ginzburg models on such spaces that
do not have known UV completions as GLSMs.  In the case of Landau-Ginzburg
models, if the metric is not Ricci-flat, not even asymptotically,
then, RG flow would be more complicated, and our analysis
likely too naive.
Our proposed hybrid Landau-Ginzburg models are constructed on the
assumption that they have Ricci-flat metrics, at least asymptotically,
so that the renormalization group flow works as expected.

It is not entirely out of the realm of possibility that complications
in RG flow, alluded to above, might actually generate nonzero $H$ flux
in a low-energy theory, especially in (0,2) supersymmetric versions of
this construction where one has less control over RG flow.  We will leave
this possibility to future work.

\section{GLSM for $\mathbb{P}^{2g+1}[2,2]$ and curves of genus $g$} 
\label{sec: map}

One essential piece of our construction will be a trick from
\cite{Caldararu:2007tc}, in which GLSMs describe geometries nonperturbatively,
rather than as the critical locus of a superpotential.  As it plays a
critical role in this paper, we review the highlights in this section.

Section $4.1$ of \cite{Caldararu:2007tc} discusses a gauged linear sigma model 
for $\mathbb{P}^{2g+1}[2,2]$ (with $g \geq 1$) which realizes 
a genus $g$ Riemann surface, via nonperturbative tricks,
in its $r \ll 0$ phase. That model will play an essential role in this
paper, so we shall quickly review it here.

The GLSM in question is a $U(1)$ gauge theory with (2,2) supersymmetry and
$2g+2$ chiral superfields $\phi_i$ of charge 1 
and two chiral superfields $p_1$, $p_2$ of charge $-2$,
with superpotential
\begin{equation}
W = p_1 Q_1(\phi) + p_2 Q_2(\phi) = \sum_{ij} \phi_i A^{ij}(p) \phi_j,  \label{eq-g-superpotential}
\end{equation}
where $Q_i$ are quadratic functions of $\phi$'s, and
$A^{ij}(p)$ is a $(2g+2) \times (2g+2)$ symmetric matrix 
whose entries are linear in the $p_a$.

For $r \gg 0$ (geometric phase), 
one can do the usual analysis of the critical loci to argue that the
GLSM flows to a sigma model on a complete intersection of two quadrics in
${\mathbb P}^{2g+1}$.

The $r \ll 0$ phase is more interesting.  D terms imply that
$p_1$ and $p_2$ can not simultaneously vanish, and the superpotential
generically gives a mass to the $\phi$'s.  On that open set where all
the $\phi$'s are massive, since the
$p$'s have nonminimal charges, physics sees a double cover of the
${\mathbb P}^1$ mapped out by $p$'s 
\cite{Caldararu:2007tc, Pantev:2005zs,Hellerman:2006zs}.
On the locus where any $\phi$ becomes massless, specifically the 
degree $2g+2$ locus
$\{ \det A = 0 \}$ where the mass matrix develops at least one zero 
eigenvalue, the double cover collapses to a single cover.

Put simply, $\{ \det A^{ij} = 0\}$ defines the branch locus on the 
double cover of $\mathbb{P}^1$.  (Monodromies about the branch locus
correspond to Berry phases and are described in
\cite{Caldararu:2007tc}.) 
The resulting geometry, a double cover of $\mathbb{P}^1$ branched over a degree $2g+2$ loci, is a compact Riemann surface of genus $g$.

\section{Hybrid Landau-Ginzburg models for fiber products}
\label{sect:genl-case}

\subsection{Fiber products with vector bundles on ${\mathbb P}^1$}
\label{sect:vec-bundle}

In this section we will consider a general set of fiber products,
between curves and vector bundles on ${\mathbb P}^1$.
Specifically, let $V$ be the total space of the
rank-two vector bundle ${\cal O}(a) + {\cal O}(b) \rightarrow 
{\mathbb P}^1$.

Mathematically, we are considering the fiber product\footnote{
As an aside, if $\pi: \Sigma \rightarrow {\mathbb P}^1$ is the projection from
the genus $g$ curve to ${\mathbb P}^1$, then the fiber product $X$ is the
total space of $L_a \oplus L_B \rightarrow \Sigma$, where
$L_n = \pi^* {\cal O}(n)$.
} 
\begin{equation*}
\begin{tikzcd}
X= \varphi^* V = V \times_{\mathbb{P}^1} \Sigma \arrow{d} \arrow{r} & V \arrow[d, "\pi"] \\
\Sigma \arrow[r,"\varphi"] & \mathbb{P}^1
\end{tikzcd}
\end{equation*}
The fiber product $X$ will have trivial canonical bundle if
\begin{equation} 
K_{\Sigma} \: = \: \varphi^*( \det V ).
\end{equation}

We will consider the special case of genus $g$ curves $\Sigma$
constructed as branched double covers, for which the condition above
reduces to
\begin{equation}   \label{eq:rk2vec:cy}
a + b \: = \: g-1.
\end{equation}

Now, in general, we will want to describes cases in which $a$ or $b$ are
positive, and the total space of such $V$ is challenging to describe with
a GLSM.  Recall that the total space of
${\cal O}(-1) \oplus {\cal O}(-1) \to \mathbb{P}^1$ can be described
by a GLSM with a single $U(1)$ gauge field and four chiral superfields:
\begin{itemize}
  \item two chiral superfields $p_a$ of charge $+1$ corresponding to homogeneous coordinates on the base $\mathbb{P}^1$,
  \item two chiral superfields $y_a$ of charge $-1$ corresponding to the two line bundles ${\cal O}(-1)$.
\end{itemize}
Naively, one could try a similar GLSM with the charges of the
chiral superfields $y_a$ flipped to $+1$ describing two ${\cal O}(+1)$
line bundles.
However, D terms in the resulting GLSM make it clear that that GLSM
will describe the space $\mathbb{P}^3$. To describe the total space of the
bundle above, one would need to remove a different exceptional locus than
the one canonically dictated by the D terms for a quotient of flat space.

Setting aside the issue above, the fiber product
would formally be described by the GLSM with 
gauge grop $U(1)$ and matter
\begin{itemize}
\item $2g+2$ chiral superfields $\phi_i$ of charge $-1$,
\item $2$ chiral superfields $p_a$ of charge $+2$,
\item $2$ chiral superfields $y_a$, $y_b$ of charges $2a$, $2b$,
\end{itemize}
and superpotential
\begin{displaymath}
W \: = \: \sum_{ij} \phi_i \phi_j A^{ij}(p),
\end{displaymath}
where $A^{ij}$ is a symmetric $(2g+2)\times(2g+2)$ matrix with entries
linear in the $p$'s.  

We have taken the matrix $A^{ij}$ to be independent of the $y$'s, to preserve
translation invariance along the fibers as well as a global $SU(2)$ rotation
symmetry between the $y$'s. 
In models discussed later we will
make similar restrictions so as to reproduce the desired geometries.

In passing, if we were to add terms to the superpotential to realize
the most general case compatible with gauge invariance,
{\it i.e.} adding terms involving $y_a$, $y_b$, and sufficient $\phi$
factors,
we would get the GLSM for ${\mathbb P}^{2g+1}[2,2,2a,2b]$
(with an overall sign flip on the charges, inverting the $r \gg 0$ and
$r \ll 0$ phases).  (This includes the GLSM for
${\mathbb P}^7[2,2,2,2]$, studied in \cite{Caldararu:2007tc} 
because of the geometric
realization of its $r \ll 0$ phase.)
Note that the Calabi-Yau condition for that complete intersection also
reduces to~(\ref{eq:rk2vec:cy}).

Assuming that $p_1$ and $p_2$ are homogeneous coordinates on $\mathbb{P}^1$,
one can easily see, modulo the issue with D terms,
that the mass matrix in the F term imply the
fiber product geometry in the phase
$r \gg 0$.
First, following the same argument in section (\ref{sec: map}),
the fields $\phi_i$ and $p_a$ describe the genus $g$ Riemann surface as the
branched double cover of $\mathbb{P}^1$. On the other hand,
the fields $y_a$ and $y_b$ correspond to the fiber coordinates 
on ${\cal O}(a) \oplus {\cal O}(b)$
since we assumed that $p_1$ and $p_2$ are homogeneous coordinates on $\mathbb{P}^1$.
The fiber product structure is achieved by identifying two different
$\mathbb{P}^1$ in the holomorphic map and the total space of $V$.
The identification is manifest in our theory.

Finally, let us consider the Calabi-Yau condition.
The sum of the charges
in this theory is precisely
\begin{displaymath}
-2g + 2 + 2a + 2b,
\end{displaymath}
and so vanishes precisely when the mathematical
condition~(\ref{eq:rk2vec:cy}) for
trivial canonical bundle is satisfied.

As mentioned above, the putative GLSM above does not quite work,
because the D terms will not describe the correct exceptional locus
in general.
To evade the issue of positive-degree line bundles on
${\mathbb P}^1$ in GLSMs, we construct a hybrid Landau-Ginzburg model,
an ungauged sigma model on the total space of
\begin{equation} \label{eq:rk2vec:lgvec}
{\cal O}(-1/2)^{2g+2} \oplus {\cal O}(a) \oplus {\cal O}(b) 
\to {\mathbb P}^1,
\end{equation}
with superpotential
\begin{equation}
W \: = \: \sum_{ij} \phi_i \phi_j A^{ij}(p),
\end{equation}
where the mass matrix $A^{ij}(p)$ should now be interpreted as a generic
symmetric $(2g+2)\times (2g+2)$ matrix of sections of
${\cal O}(+1) \to {\mathbb P}^1$.

Notice that the $\mathbb{P}^1$ in the target space~(\ref{eq:rk2vec:lgvec}) 
is actually a $\mathbb{Z}_2$ gerbe on $\mathbb{P}^1$ indicated by the
line bundle denoted $\mathcal{O}(-1/2)$. The bundle $\mathcal{O}(-1/2)$ is a 
special kind of line bundle that only exist for gerbes. On the other hand,
it is also a fiber bundle of $\mathbb{P}^1$ whose fibers are the orbifolds
$[\mathbb{C}/\mathbb{Z}_2]$.
More details of such line bundles on gerbes over projective spaces are
discussed in appendix B of \cite{Anderson:2013sia}.

Generically on the ${\mathbb P}^1$, the $\phi_i$ are massive,
away from the locus $\{ \det A = 0 \}$, and so the ${\mathbb Z}_2$
gerbe implies a branched double cover, as usual, and hence the fiber
product of the genus $g$ Riemann surface and the vector bundle $V$
over ${\mathbb P}^1$.

The condition for the canonical class to be trivial in this hybrid model
is that the first Chern class of the vector bundle~(\ref{eq:rk2vec:lgvec})
match the first Chern class of the canonical bundle, meaning specifically
that
\begin{displaymath}
(2g+2)(-1/2) + a + b \: = \: -2,
\end{displaymath}
which again reduces to the mathematical condition~(\ref{eq:rk2vec:cy}).

There is a possible technical issue with this construction, due to the
fact that there is no non-compact version of Yau's theorem, as described
in the introduction.  We describe
above a hybrid Landau-Ginzburg model over a K\"ahler space with holomorphically
trivial canonical bundle (in the case $g=3$); however, that does not
guarantee that a Ricci-flat metric exists in the noncompact case.
If the metric is not Ricci-flat, at least asymptotically, then the RG flow
may be more complicated than we have naively supposed. 
If the model arose from a GLSM, we could appeal to RG flow from the GLSM,
but as we have not been able to write down a UV GLSM, we cannot guarantee
that an asymptotically Ricci-flat metric exists.  Analogous potential
issues arise
in every hybrid Landau-Ginzburg model described in this paper.

\subsection{Fiber products with hypersurfaces in vector bundles}
\label{sect:genl-hyp}

Let $V$ denote the rank-three vector bundle
\begin{displaymath}
{\cal O}(a) \oplus {\cal O}(b) \oplus {\cal O}(c) \to
{\mathbb P}^1,
\end{displaymath}
and consider the hypersurface $f(x,y,z)=0$, where $x$, $y$, $z$ are
coordinates along the fibers of $V$, and $f$ is of degree\footnote{
In the sense of weighted projective spaces, so that the monomials
$x$, $y$, $z$ have weights $a$, $b$, $c$, respectively.
} $d$.  Mathematically, for the fiber product of
this hypersurface with the curve $\Sigma$ of genus $g$
to have trivial canonical
bundle, we must require that the degree $d$ match the degree of $\varphi^* V$,
for $\varphi: \Sigma \rightarrow {\mathbb P}^1$, which means,
for $\Sigma$ realized as a branched double cover of ${\mathbb P}^1$,
\begin{equation}
d \: = \: a + b + c + 1 - g.
\end{equation}

Modulo the same issue with D terms and positive-degree line bundles
discussed in the last section,
we can construct a `fake' GLSM for the fiber product with the hypersurface
in $V$ as a $U(1)$ gauge theory with matter
\begin{itemize}
\item $2g+2$ chiral superfields $\phi_i$ of charge $-1$,
\item $2$ chiral superfields $p_a$ of charge $+2$,
\item $3$ chiral superfields $x$, $y$, $z$ of charges $2a$, $2b$, $2c$,
respectively,
\item $1$ chiral superfield $q$ of charge $-2d$,
\end{itemize}
and superpotential
\begin{displaymath}
W \: = \: \sum_{ij} \phi_i \phi_j A^{ij}(p)
\: + \:
q f(x,y,z),
\end{displaymath}
where $A^{ij}$ is a symmetric $(2g+2)\times(2g+2)$ matrix with entries
linear in the $p$'s.  (As before, we do not consider more general possible
terms, in order to preserve pertinent symmetries.) 
The sum of the charges in this theory vanishes when
\begin{equation}  \label{eq:fib-hyp:cy}
g + d \: = \: a + b + c + 1,
\end{equation}
matching the mathematical condition given above for the canonical
bundle of the fiber product to be trivial.

As before, there is a problem involving D terms in the putative GLSM
above.  To evade this issue,
we can construct a hybrid Landau-Ginzburg model which describes
the geometry.  Specifically, this will be an ungauged sigma model on the
total space of
\begin{equation}   \label{eq:lg:fib-hyp}
{\cal O}(-1/2)^{2g+2} \oplus {\cal O}(a) \oplus {\cal O}(b) \oplus
{\cal O}(c) {\cal O}(-d) \to {\mathbb P}^1,
\end{equation}
where we interpret the bundle in terms of a ${\mathbb Z}_2$ gerbe
on the ${\mathbb P}^1$,
and with superpotential
\begin{displaymath}
W \: = \: \sum_{ij} \phi_i \phi_j A^{ij}(p)
\: + \:
q f(x,y,z),
\end{displaymath}
where the mass matrix $A^{ij}(p)$ is a generic symmetric $(2g+2) \times (2g+2)$
matrix of sections ${\cal O}(+1) \to {\mathbb P}^1$.

This hybrid Landau-Ginzburg model realizes the same fiber product structure
as the fake GLSM above.  The superpotential contains a mass matrix for the
$\phi_i$, $i=1,\ldots, 2g+2$, that gives them a mass away from the locus
$\{ \det A = 0 \}$.  As a result, at generic points on the ${\mathbb P}^1$,
the remaining massless fields are invariant under the gerbe ${\mathbb Z}_2$,
which physics sees \cite{Caldararu:2007tc,Hellerman:2006zs}
as a double cover of ${\mathbb P}^1$, branched over the locus
$\{ \det A = 0 \}$.  Consequently, one obtains a fiber product of the genus
$g$ curve and the hypersurface.

The Calabi-Yau condition for the hybrid Landau-Ginzburg model is the
condition that $c_1$ of the bundle~(\ref{eq:lg:fib-hyp}) match $c_1$ of
the canonical bundle of ${\mathbb P}^1$, which in this case implies
\begin{displaymath}
(2g+2)(-1/2) + a + b + c -d \: = \: -2.
\end{displaymath}
It is straightforward to check that this matches the 
condition~(\ref{eq:fib-hyp:cy}) given earlier.

\section{Fiber products with twistor spaces} \label{sec:hybrid}

In this section, we will construct (2,2) supersymmetric hybrid Landau-Ginzburg
theories which should RG flow to sigma models on the non-compact K\"ahler
Calabi-Yau threefolds constructed as fiber products of genus three
curves and twistor
spaces, as explained in the introduction.
The (K\"ahler) twistor spaces we consider here are the twistor spaces\footnote{
Sometimes, blowdowns of the twistor spaces.
} 
of $\mathbb{R}^4$,  $\mathbb{C}^2 / \mathbb{Z}_k$ 
and $S^1 \times \mathbb{R}^3$.   These will all correspond to
special cases of the constructions in section~\ref{sect:genl-case},
so we will strive to be brief.
In each case, since the curve is realized as a branched double
cover of ${\mathbb P}^1$, the Calabi-Yau condition is that the curve
be of genus 3 -- details of the hyperK\"ahler manifold are otherwise
irrelevant.  In each of our models, we will recover the genus three condition
as a consistency check.

\subsection{Fiber product with twistor space of $\mathbb{R}^4$} \label{sec:r4}

From \cite{Hitchin:1983}, the twistor space of $\mathbb{R}^4$ can be described 
as the total space of 
$\mathcal{O}(+1) \oplus \mathcal{O}(+1) \to \mathbb{P}^1$. 
Our construction for this case is a special case of the
construction in section~\ref{sect:vec-bundle}.

As discussed in section~\ref{sect:vec-bundle}, there is a technical question
of how to realize positive-degree line bundles in GLSMs, so we 
instead construct a lower energy theory, a hybrid Landau-Ginzburg model. 
Specifically, this will be an ungauged sigma model on the total space of 
\begin{equation} 
\mathcal{O}(-1/2)^{2g+2} \oplus \mathcal{O}(+1) \oplus \mathcal{O}(+1) \to \mathbb{P}^1, \label{totalspace}
\end{equation}
with superpotential 
\begin{equation*}
W = \sum_{ij} \phi_i \phi_j A^{ij}(p),
\end{equation*}
where 
the mass matrix
$A^{ij}(p)$ should now be interpreted as a generic symmetric $(2g+2)\times
(2g+2)$ matrix of
sections of ${\cal O}(+1) \rightarrow {\mathbb P}^1$.

The superpotential contains a mass matrix for the $\phi_i, i=1,\dots,2g+2$, 
that gives them a mass away from the locus $\{\det A = 0\}$. 
Therefore, at generic points on the $\mathbb{P}^1$, 
the remaining massless fields are all non-minimally charged. 
The Riemann surface of genus $g$ is given by a double cover of $\mathbb{P}^1$ 
branched over a degree $2g+2$ locus as before. 
Also, the fields $y_1$ and $y_2$ are the coordinates on the fibers 
of $\mathcal{O}(+1) \oplus \mathcal{O}(+1)$ of the same $\mathbb{P}^1$. 
Consequently, one obtains a fiber product of the genus $g$ Riemann surface 
and twistor space of $\mathbb{R}^4$ over ${\mathbb P}^1$. 

The Calabi-Yau condition for the total space of a vector bundle over 
$\mathbb{P}^1$ is that the first Chern class of the vector bundle should 
be the same as the first Chern class of the canonical bundle
of ${\mathbb P}^1$. In this case, one gets
\begin{equation*}
(2g+2)(-1/2) +1+1 = -2.
\end{equation*}
It implies that the genus of the Riemann surface is three, as expected.

\subsection{Fiber product with twistor space of $\mathbb{C}^2 / \mathbb{Z}_k$}  \label{sec:zk}

Next, we will give a physical theory describing the fiber product with 
(the blowdown of) a 
different twistor space, namely the
the twistor space of $\mathbb{C}^2 / \mathbb{Z}_k$ 
\cite{Hitchin:1983, Hitchin:1979}. 
The group $\Gamma = {\mathbb Z}_k$ 
acts on $\mathbb{C}^2$ as follows:
\begin{equation*}
(z_1,z_2) \to (e^{2 \pi i n/k} z_1, e^{-2 \pi i n/k} z_2).
\end{equation*}
Notice that the monomials $x = z_1^k$, $y = z_2^k$, $z = z_1 z_2$ are invariant under the group $\Gamma$. Therefore the singular surface $\mathbb{C}^2 / \Gamma$ can be described by a hypersurface embedding in $\mathbb{C}^3$,
\begin{equation*}
\{ x y = z^k \} \subset \mathbb{C}^3 = \text{Spec} \, \mathbb{C}[x,y,z].
\end{equation*}
One can turn on a universal family of complex structure deformations 
which is given by adding lower order terms in $z$. 
As a result, the hypersurface defining equation becomes 
\begin{equation*}
x y = z^k + a_1 z^{k-1} + \dots + a_k = \prod_{i=1}^k (z - f_i),
\end{equation*}
where $a_i$ and $f_i$ are constant parameters. 
The twistor space is a resolution of the
hypersuface 
\begin{equation*}
\{x y = \prod_{i=1}^k (z - f_i (p))   \} \subset \text{Tot} (\mathcal{O}(+k) \oplus \mathcal{O}(+k) \oplus \mathcal{O}(+2) \to \mathbb{P}^1),
\end{equation*}
where $x$, $y$ are fiber coordinates on the bundle $\mathcal{O}(+k)$, $z$ 
on $\mathcal{O}(+2)$, and $f_i$ are sections of 
$\mathcal{O}(+2) \to \mathbb{P}^1$. 
In particular, for each point of $\mathbb{P}^1$, 
the fiber is a deformation of $\mathbb{C}^2 / \mathbb{Z}_k$. 
In the special case $k=2$, the fiber space is also known as 
an Eguchi-Hanson space. 

We can realize the hypersurface above, a blowdown of the twistor space,
using the same ideas as in section~\ref{sect:genl-hyp}.
Specifically, we propose a hybrid Landau-Ginzburg model, an
ungauged sigma model whose target space is the total space of 
\begin{equation*}
\mathcal{O}(-1/2)^{2g+2} \oplus \mathcal{O}(+k)^2 \oplus \mathcal{O}(+2)\oplus \mathcal{O}(-2k)  \to \mathbb{P}^1,
\end{equation*}
with fiber coordinates $\phi_i$ on $\mathcal{O}(-1/2)^{2g+2}$, $x$, $y$ on $\mathcal{O}(+k)^2$, $z$ on $\mathcal{O}(+2)$ and $q$ on $\mathcal{O}(-2k)$. 
The superpotential is 
\begin{equation}
W =  \sum_{ij} \phi_i \phi_j A^{ij}(p) + q (x y - \prod_{i=1}^k (z - f_i (p)) ), \label{eq:sp4}
\end{equation}
where $f_i$ are sections of ${\cal O}(2) \rightarrow {\mathbb P}^1$
and
$A^{ij}(p)$ is symmetric $(2g+2)\times (2g+2)$ matrix with elements which
are sections of ${\cal O}(1) \rightarrow {\mathbb P}^1$.

Going through the same analysis as in section~\ref{sect:genl-hyp},
one obtains the desired fiber product geometry. Note that the Calabi-Yau condition in this case is 
\begin{equation*}
(−1/2)(2g + 2) + k+k+2 + (−2k) = - 2
\end{equation*}
hence $g=3$, as expected.

\subsection{Fiber product with twistor space of $S^1 \times \mathbb{R}^3$}

The last case we will discuss here is the fiber product with
(the blowdown of)
the twistor space of $S^1 \times \mathbb{R}^3$ \cite{Hitchin:1983}. 
Since the analyses is similar to previous sections, 
we will present our proposition briefly here. 
The space $S^1 \times \mathbb{R}^3$ can be defined as $\mathbb{C}^2 / \Gamma$ where $\Gamma $ is given by
\begin{equation*}
(z_1, z_2) \to (z_1 + 1, z_2).
\end{equation*}
Following the same process as in section~\ref{sec:zk}, 
the twistor space is defined by a resolution of the hypersurface 
\begin{align*}
\{y^2 + z x^2 =& z + (f_1(p)^2 +f_2(p)^2) + 2 f_1(p)^2  f_2(p)^2 x   \}  \\
& \quad \quad \quad \quad \subset 
\text{Tot} \, (\mathcal{O} \oplus \mathcal{O}(+2) \oplus \mathcal{O}(+4) \to \mathbb{P}^1),
\end{align*}
where $x$ is a fiber coordinate on the line bundle $\mathcal{O}$, 
$y$ on $\mathcal{O}(+2)$, $z$ on $\mathcal{O}(+4)$ 
and $f_1$, $f_2$ are two sections of $\mathcal{O}(+2) \to \mathbb{P}^1$.

We can construct a hybrid Landau-Ginzburg model realizing this geometry
as a special case of the construction in section~\ref{sect:genl-hyp}.
This hybrid Landau-Ginzburg model is defined on the total space of 
\begin{equation*}
\mathcal{O}(-1/2)^{2g+2} \oplus \mathcal{O}(0)\oplus \mathcal{O}(+2)\oplus \mathcal{O}(+4) \oplus \mathcal{O}(-4) \to \mathbb{P}^1,
\end{equation*}
with superpotential  
\begin{equation}
W =  \sum_{ij} \phi_i \phi_j A^{ij}(p) + q (y^2 + z x^2 - z - (f_1(p)^2 +f_2(p)^2) - 2 f_1(p)^2  f_2(p)^2 x ), \label{eq:sp3}
\end{equation}
where $A^{ij}$ is a symmetric $(2g+2)\times(2g+2)$ matrix with entries
that are sections of ${\cal O}(1) \rightarrow {\mathbb P}^1$.

The Calabi-Yau condition 
\begin{equation*}
(−1/2)(2g + 2) + 3(+2) + (−4) = −2
\end{equation*}
implies that $g=3$, as expected.

\section{Conclusions}

In this paper, we have constructed hybrid Landau-Ginzburg models that
RG flow to 
a new family of non-compact Calabi-Yau threefolds, constructed
as fiber products of genus $g$ curves and 
noncompact K"ahler threefolds.
We only consider curves given as branched double covers of $\mathbb{P}^1$.
Our construction utilizes `nonperturbative' constructions of the
genus $g$ curves given in \cite{Caldararu:2007tc}, and so provides
a new set of exotic UV theories that should RG flow to sigma models
on Calabi-Yau manifolds, in which the Calabi-Yau is not realized simply
as the critical locus of a superpotential.

As important special cases,
we applied these constructions to describe fiber products with
certain K\"ahler twistor spaces
of noncompact hyperK\"ahler four-manifolds, specifically
$\mathbb{R}^4$, $\mathbb{C}^2 / \mathbb{Z}_k$ and $S^1 \times \mathbb{R}^3$. 
We check that the Calabi-Yau condition one sees in physics matches that
from mathematics, namely that the curve have genus three, independent of
details of the four-manifold.

We see this work as a first step to realizing GLSMs for compact non-K\"ahler
analogues of Calabi-Yau threefolds constructed in \cite{Fei:2017ctw},
to which we hope to return in the future.

\section{Acknowledgements}

We would like to thank L.~Anderson, T.~Fei, J.~Gray, P.~Gao, and I.~Melnikov 
for useful conversations, and in particular P.~Gao for the question
that led to this paper. 
E.S. was partially supported by NSF grant PHY-1720321.
T.P. was supported in part by NSF grant DMS 1601438, 
by Simons HMS Collaboration grant grant \# 347070, 
and through a research visit to the Laboratory of Mirror Symmetry 
NRU HSE supported by a  RF Government grant, ag. No 14.641.31.0001.

\appendix

\section{Review of pertinent mathematics} \label{sec:math}

According to proposition 2.2 of \cite{Fei:2017ctw},
the fiber product of a twistor space $X$
and a genus $g$ curve $\Sigma$ with {\it some} map
$\varphi: \Sigma \rightarrow {\mathbb P}^1$ has trivial
canonical bundle if and only if 
\begin{equation}  \label{eq:triv-canonical-class}
\varphi^* {\cal O}(2) \cong K_{\Sigma}.
\end{equation}

We can see this as follows.
Let $\pi: X \rightarrow {\mathbb P}^1$ be the twistor space for
any hyperK\"ahler surface,
then the relative symplectic form is a nowhere-zero section
of $K_{X/\mathbb{P}^{1}} \otimes \pi^{*}\mathcal{O}(2)$,
hence $K_{X/\mathbb{P}^1} \cong \pi^* {\cal O}(-2)$. 
Furthermore, by definition,
\begin{eqnarray*}
K_{X/\mathbb{P}^1} & = & K_X \otimes \pi^* K_{\mathbb{P}^1}^{-1}, \\
& = & K_X \otimes \pi^* {\cal O}(2).
\end{eqnarray*}
This gives
\begin{displaymath}
K_{X} = \pi^{*}\mathcal{O}(-4).
\end{displaymath}
Next, for the fiber product $Z = X \times_{ {\mathbb P}^1 } \Sigma$,
\begin{eqnarray*}
K_{Z/\Sigma} & = & p_X^* K_{X/{\mathbb P}^1}, \\
& = & p_X^* \pi^* {\cal O}(-2) = \pi_Z^* {\cal O}(-2),
\end{eqnarray*}
using
$\pi_Z = \pi \circ p_X = \varphi \circ p_{\Sigma}$, and where
$p_{\Sigma}: Z \rightarrow \Sigma,$ $\pi_Z: Z \rightarrow 
{\mathbb P}^1$ are projections.
Hence,
\begin{eqnarray*}
K_{Z} & = & p_{\Sigma}^* K_{\Sigma} \otimes K_{Z/\Sigma}, \\
& = & p_{\Sigma}^{*}K_{\Sigma} \otimes \pi_Z^{*} \mathcal{O}(-2).
\end{eqnarray*}
To double-check, we can also compute
\begin{eqnarray*}
K_Z & = & p_X^* K_X \otimes K_{Z/X}, \\
& = & p_X^* K_X \otimes p_{\Sigma}^* K_{\Sigma/\mathbb{P}^1}, \\
& = & p_X^* \pi^* {\cal O}(-4) \otimes p_{\Sigma}^* \left( K_{\Sigma} \otimes
\varphi^* {\cal O}(2) \right), \\
& = & \pi_Z^* {\cal O}(-4) \otimes p_{\Sigma}^* K_{\Sigma} \otimes
\pi_Z^* {\cal O}(2), \\
& = & p_{\Sigma}^* K_{\Sigma} \otimes \pi_Z^* {\cal O}(-2),
\end{eqnarray*}
matching the result above.
In any event, 
using the fact that $\pi_Z = \pi \circ p_X = \varphi \circ p_{\Sigma},$
we see that $K_Z$ is trivial if and only if 
\begin{displaymath}
K_{\Sigma}  = \varphi^* {\cal O}(2).
\end{displaymath}

The fact that this condition does not depend upon $X$ follows from the
fact that $K_X$ is always a pullback of ${\cal O}(-4)$.  

Now, under what circumstances is~(\ref{eq:triv-canonical-class})
satisfied?

Let us consider the case that $\Sigma$ is a spectral cover of 
${\mathbb P}^1$, the case of relevance for this paper.
Let $d$ be the degree of the projection map $\varphi: \Sigma 
\rightarrow {\mathbb P}^1$, then $2d = 2g-2$ for $g$ the genus of
$\Sigma$, hence $g = d+1$.  From Hurwitz,
$K_{\Sigma} = \varphi^{*}\mathcal{O}(-2)\otimes
\mathcal{O}(R),$ where
$R \subset \Sigma$ is the ramification divisor of $\varphi: \Sigma
\rightarrow {\mathbb P}^1$.  Thus, to satisfy~(\ref{eq:triv-canonical-class}),
we must satisfy
$ \mathcal{O}(R) =
\varphi^{*}\mathcal{O}(4).$

For a hyperelliptic double cover, the degree of the ramification divisor
is $2g+2$ and $\mathcal{O}(R) = f^{*}\mathcal{O}(g+1).$
Thus, in such a case, one must require $g=3$.

More generally spectral covers have the property that their
ramification divisor is a pullback. If $\varphi : \Sigma \to \mathbb{P}^{1}$
is a spectral cover of degree $d$ emebedded in the total space of
$\mathcal{O}(k)$, then $\mathcal{O}(R) = \varphi^{*}\mathcal{O}((d-1)k).$
To satisfy~(\ref{eq:triv-canonical-class}), one must require
$(d-1)k =4,$ which gives three options (for the case that
$\Sigma$ is a spectral cover of ${\mathbb P}^1$):
\begin{itemize}
\item $d-1 = 1$, $k = 4,$ which gives a hyperelliptic curve of genus 3
(the case that arises in this paper),
\item $d-1 = 2$, $k = 2,$ which gives a 3-sheeted spectral cover
of ${\mathbb P}^1$ of genus 4,
\item $d-1 = 4$, $k = 1,$ which gives a 5-sheeted spectral cover
of ${\mathbb P}^1$ of genus 6.
\end{itemize}
Note this condition is not satisfied for $\Sigma$ a genus one curve.

If we drop the spectral curve constraint on $\Sigma$, 
then there are solutions in any genus $\geq 3$.
We can see this as follows.  First, the 
condition~(\ref{eq:triv-canonical-class}) for the fiber product to have
trivial canonical class can be rephrased as the statement that 
$\varphi$ is given by a basepoint-free pencil of sections in some 
spin structure $L$ on $\Sigma$.  
From a theorem of Harris \cite{harris}, if $g \geq 3$,
the moduli space of pairs $(\Sigma, L)$ such that $\Sigma$ is smooth of
genus $g$ and $L$ is a spin structure which has a pencil of sections
has dimension $3g-4$.

For another example, there is a theorem of Farkas \cite{farkas}
which says
that for any $g \geq 10$, the moduli space of pairs $(\Sigma, L)$
such that $\Sigma$ is a smooth curve of genus $g$ and
$L$ is a spin structure which gives an embedding of $\Sigma$ in
${\mathbb P}^3$ is of dimension $3g-9$.  Given such a pair and compose
the embedding $\phi_L: \Sigma \rightarrow {\mathbb P}^3$ with a generic
projection from some point not in the image, one will get a morphism
$\Sigma \rightarrow {\mathbb P}^1$ which has the desired property.
Thus, for $g \geq 10$, there is a $3g-9$-dimensional space of pairs
with the desired property.

So far we have discussed conditions for the fiber product $Z$ to have
holomorphically trivial canonical bundle.  Next, let us turn to the
question of when $Z$ is K\"ahler.  Since $Z$ is a finite cover of $X$,
$Z$ is K\"ahler if and only if $X$ is K\"ahler.

\end{document}